\documentclass[prb,twocolumn,showpacs,preprintnumbers,amsmath,amssymb]{revtex4}

\usepackage{graphicx}
\usepackage{psfig}
\usepackage{dcolumn}
\usepackage{bm}

\begin{document}
\newcommand{\etal}{{\it etal.\/}}
\newcommand{\degc}{{$^{o}$}}

\title{
Critical properties 
in single crystals of Pr$_{1-x}$Pb$_{x}$MnO$_{3}$
}

\author{B. Padmanabhan$^{a}$,\footnote{
corresponding author. fax: 91-80-360 2602, 
E-mail: paddu@physics.iisc.ernet.in}
 H. L. Bhat$^{a}$, Suja Elizabeth$^{a}$
}

\affiliation{$^{a}$Department of Physics, Indian Institute
of Science,\\ Bangalore 560012, India}

\author{Sahana R\"o{\ss}ler$^{b,c,d}$,
U.K. R\"o{\ss}ler$^b$,
K.D\"orr$^b$,
K. H.M\"uller$^{b}$}
\affiliation{$^{b}$IFW Dresden, Postfach 270016, D-01171 Dresden, Germany}
\affiliation{$^{c}$Institut f\"ur Festk\"{o}rperphysik, 
Technische Universit\"{a}t Dresden, D-01062 Dresden, Germany}
\affiliation{$^{d}$Max-Planck-Institut f\"ur Chemische Physik fester Stoffe,
N\"othnitzer Str. 40, D-01187 Dresden, Germany}

\date{\today}

\begin{abstract}
The critical properties at the ferromagnetic - 
paramagnetic transition have been analysed 
from data of static magnetization measurements 
on single crystals of Pr$_{1-x}$Pb$_{x}$MnO$_{3}$, for $x=$~0.23 
and $x=$~0.30. 
In Pr$_{1-x}$Pb$_{x}$MnO$_{3}$ 
the ferromagnetic ordering and the metal-insulator transition 
do not coincide in parts of the phase diagram.
The crystal with $x=$~0.23 is a ferromagnetic
insulator with Curie temperature $T_{\textrm{C}}=$~173~K, 
while the crystal with $x=$0.30 has 
 $T_{\textrm{C}}=$~198~K and remains 
metallic up to a metal-insulator transition temperature 
$T_{\textrm{MI}}=$~235~K.
The dc magnetization measurements were carried 
out in the field range from 0 to 5 T for an interval 
in the critical temperature range  
$T_{\textrm{C}} {\pm}$10~K corresponding to a reduced temperature interval
$ 0.003 < \epsilon < 0.6 $.
The exponents ${\beta}$ for spontaneous magnetization, 
${\gamma}$ for the initial 
susceptibility above $T_{\textrm{C}}$ 
and ${\delta}$ for the critical magnetization isotherm 
at $T_{\textrm{C}}$
were obtained by static scaling analysis from modified Arrott 
plots and by the Kouvel Fisher method
for the insulating crystal with composition $x=$~0.23.
The data are well described by critical exponents 
similar to those expected for the Heisenberg 
universality class relevant for conventional isotropic magnets 
Systematic deviations from scaling in the 
data for the metallic crystal with composition $x=0.30$ 
are demonstrated from effective critical exponents 
near the assumed ordering transition.
The unconventional magnetic ordering in this system 
indicates the presence of frustrated magnetic couplings
that suppresses magnetic ordering and lowers the 
transition temperature.
\end{abstract}

\noindent 

\maketitle

\section{Introduction}\label{Intro}
The research in rare earth manganites started around fifty years ago. 
The simultaneous occurrence of ferromagnetism and 
metallicity in these compounds 
has been explained by Zener's double exchange (DE) 
mechanism.\cite{Zener51}
%
According to this mechanism in mixed-valence manganites 
 the hopping of e$_{g}$ electron from Mn$^{3+}$ to Mn$^{4+}$ 
ion mediated by the intervening 2p level of the O$^{2-}$ ion
leads to a ferromagnetic alignment of the spins 
in the core-like t$_{2g}$ states.
In recent years there has been a revived interest 
in manganites because of the large colossal magnetoresistance (CMR) 
effect shown by these materials, 
which is caused by the suppression of spin scattering 
around the ferromagnetic transition temperature.
\cite{Tokura00,CNRRao98}
%
The maximum CMR in manganites 
is shown by compositions which fall in the 
range $x = 0.30$ to $x = 0.45$.\cite{CNRRao98}
La$_{1-x}$Sr$_{x}$MnO$_{3}$ which is the canonical double exchange 
manganite with maximum metallicity and ferromagnetism 
has a paramagnetic--ferromagnetic (PM-FM) ordering temperature 
up to about $T_{\textrm{C}}=350$~K in this composition range.
But it shows only relatively 
weak magnetoresistance.\cite{Urushibara95}
On the other hand, La$_{1-x}$Ca$_{x}$MnO$_{3}$ has 
a low $T_{\textrm{C}} \simeq 250$~K at optimal doping
and shows a sharp metal-insulator (MI) 
transition at a similar temperature $T_{\textrm{MI}}$ 
with an associated large magnetoresistivity 
close to these temperatures.\cite{Schiffer96,Martin99}

The PM--FM transition accompanied by the MI transition is seen 
as a crossover from localisation to delocalisation of charge carriers. 
Thus to elucidate the cause of CMR effect near the MI transition, 
one needs to understand the precise nature of the FM--PM transition. 
Due to the Jahn--Teller (JT) effect in the Mn$^{3+}$ ions,
one assumes that strong couplings exist between 
the itinerant e$_{g}$ electron and JT phonons.
This coupling to the lattice can result in the formation of polarons. 
Hence the MI transition can be described as 
a discontinuous transition between the mobile 
and localised polarons. 
As the mobile carriers are strongly coupled to the t$_{2g}$ spins,
this mechanism, extended by JT lattice couplings, can lead
to the formation of spin-polarized polarons.
This mechanism explains the link between an
unconventional FM--PM transition and the 
MI transition by mobile spin polarons.
Since this mechanism involves a discontinuous formation 
of polarons, 
Goodenough {\it et al}.\cite{Archibald96}
suggested that the FM--PM 
transition should be of first order.
Contrary to this, most of the manganites 
show a second order phase transition at 
the Curie temperature $T_{C}$. 
The usual perception that the ferromagnetic 
transitions in the mixed-valence manganites 
are of second order is corroborated by the absence of hysteresis 
in the temperature variation of magnetization and the observation 
of a smaller entropy change 
from specific heat data across the transition. 
Theoretical investigations on the critical 
behavior of CMR manganites by Motome and Furukawa\cite{Motome01}
based 
on simplifed DE models reveal that the FM--PM transition 
should belong to the Heisenberg universality class. 
The Heisenberg model assumes short-range interactions 
between localised spins. 
But, the range of exchange interactions
in CMR manganites is not obvious due to the itinerant 
nature of the e$_{g}$ electrons and the close link between 
 metal like conductivity and ferromagnetism.
Therefore, it is not clear whether the ferromagnetic
manganites should generically behave like conventional
isotropic magnets with a FM--PM transition belonging
to the Heisenberg universality class.

From experiments, varied and in part contradictory 
results have been obtained using different measurement 
techniques and data analyses to determine the critical properties. 
In one of the earlier papers, Lofland {\it et al.}\cite{Lofland97}
using microwave absorption technique 
on La$_{0.7}$Sr$_{0.3}$MnO$_{3}$, 
have estimated ${\beta}$ = 0.45, 
which is close to the mean field exponent $\beta=0.5$.
However, Martin {\it et al.}\cite{Martin96} 
report ${\beta}$ = 0.295 for the same system 
from neutron scattering data,
which is closer to the value of the exponent for 
the Ising model $\beta=0.325$.
In La$_{1-x}$Ca$_{x}$MnO$_{3}$, the nature of 
the transition is found to be extremely sensitive 
to the divalent doping concentration.
For example, in composition $x$ = 0.2 which is 
a ferromagnetic insulator, 
a continuous transition is observed.\cite{Hong01}
In  ferromagnetic metallic La$_{1-x}$Ca$_{x}$MnO$_{3}$ 
for $x= 0.33$ 
a first order phase transition is 
observed, while a tricritical behavior has been reported 
for $x=0.4$.\cite{Kim02}
The physical properties of Pb doped PrMnO$_{3}$ differ from those
of the La-based manganites, 
since Pr$^{3+}$ has a lower ionic radius than La. 
Hence the system has 
a reduced average radius of the A-site ions, $\left<r_{A}\right>$, 
and a lower $T_{\textrm{C}}$. 
Interestingly, in a certain doping range, 
this manganite system shows a rather wide temperature 
interval with a paramagnetic metallic state 
above $T_{\textrm{C}}$ and below the maximum in 
the resistivity curve $\rho(T)$ at $T_{\textrm{MI}}$ 
that signals the metal-insulator transition.\cite{Padmanabhan06}
Hence, depending on the doping level, 
one finds in the phase diagram for Pr$_{1-x}$Pb$_{x}$MnO$_{3}$ 
a PM--FM transition from a metallic paramagnetic phase 
into a ferromagnetic phase. 
This feature is unusual because 
in the majority of the mixed-valent
ferromagnetic manganites one finds $T_{\textrm{C}} \simeq T_{\textrm{MI}}$.
Indeed, assuming that the DE mechanism is responsible 
for the ferromagnetic order, it is difficult to explain
a wide intermediate paramagnetic phase with metal-like conductivity. 
At lower doping one finds a transition 
between an insulating paramagnetic phase 
into a ferromagnetic insulating phase. 
However in the vicinity of $T_{C}$ 
a metal-like conductivity behavior 
is observed. 
In a simple picture, the presence of 
a band of mobile carriers should mediate 
a sufficiently strong ferromagnetic exchange to 
drive a ferromagnetic ordering transition. 
On the other hand, the presence of PM--FM transitions 
in mixed-valent manganites without a metal-like conductivity 
is well-known and mechanisms to explain such phases 
have been proposed.\cite{Pai03}
Here, localization of carriers due to strong electron-lattice
coupling via Jahn-Teller effects or disorder may prevent
metallic conductivity, while the DE-exchange mechanism 
remains sufficiently strong to ensure a PM--FM transition.
In view of the unusual phase diagram of 
the Pr$_{1-x}$Pb$_{x}$MnO$_{3}$ system,
an investigation on the critical 
point phenomena at its PM--FM transition
and a comparison between the insulating
and the metallic ferromagnets from this system 
is of high interest to disentangle the 
mechanisms leading to ferromagnetic order and 
to MI transitions in the mixed-valent manganites.

\section{Methods}\label{Meth}
\subsection{Experimental Details}\label{Exps}
Single crystals of Pr$_{1-x}$Pb$_{x}$MnO$_{3}$ 
were grown by the flux technique.\cite{Morrish69}
The phase was confirmed by powder X-ray diffraction 
and compositions were determined by energy dispersive X-ray 
analysis (EDAX) and subsequently by inductively coupled 
plasma atomic emission spectroscopy(ICPAES) for higher precision. 
From the grown crystals, two compositions, $x=$ 0.23 and $x=$ 0.30
were selected for the present study. Basic physical properties of the
crystals from this compound series have been determined
earlier, as reported in Ref 13.
The crystal with composition, $x=$ 0.23 is a  ferromagnetic insulator 
with a Curie temperature $T_{\textrm{C}}$ of about 173 K, 
which was estimated from temperature 
variation of magnetization carried out at 0.01~T.\cite{Padmanabhan06}

The $x=$ 0.30 composition has $T_{\textrm{C}}$ = 200 K 
and it undergoes an MI transition at 235 K 
with a peak magnetoresistance 
value of 93\% near $T_{\textrm{MI}}$.\cite{Padmanabhan06}
For the present studies dc magnetization measurements $M$ vs. $H$ 
were carried out in the temperature range 155 to 179 K 
for $x=$  0.23 and from 188 to 214.5 K for $x=$ 0.30 in external fields
up to 5~T. 
The magnetization measurements were carried out 
by using commercial SQUID magnetometer (Quantum Design). 
%
The magnetic field was applied along the long edge 
of the platelet shaped crystals to minimize demagnetization effects. 
The orientation of these crystals has been 
determined by X-ray diffraction. It shows that 
the magnetic field has been oriented 
within a few degrees along a $<$100$>$ direction 
of the pseudocubic pervoskite crystal structure of 
our crystals. The effective internal field $H_{\textrm{eff}}$ 
used for the scaling analysis 
has been corrected for demagnetization,
$H_{\textrm{eff}} = H_{\textrm{appl}} - DM$, 
where D is the demagnetization 
factor obtained from $M vs. H$ 
measurements in the low-field linear response regime at 10 K. 
In addition, specific heat measurements through 
the PM--FM transition were carried out by 
a PPMS (Quantum Design). 
The detailed analysis of specific heat studies 
on these crystals will be published elsewhere.

\subsection{Scaling analysis}\label{Scaling}
The critical properties of a magnetic system
showing a second order phase transition 
are characterized by critical 
exponents ${\alpha}$, ${\beta}$, ${\gamma}$ and ${\delta}$ 
and the magnetic equation of state which connects the four exponents. 
The mathematical definitions of the exponents are given below:
The specific heat near $T_{\textrm{C}}$ is given by,
\newline
\begin{equation}
C_{+}(T) = A({\epsilon})^{-\alpha}/{\alpha}, {\epsilon} {>} 0
\end{equation}
\begin{equation}
C_{-}(T) = A'({\epsilon})^{-\alpha'}/{\alpha'},{\epsilon} {<} 0
\end{equation}
where $A$ and $A'$ are the critical amplitudes 
below and above $T_{\textrm{C}}$ respectively, 
and ${\epsilon} = (T-T_{\textrm{C}})/T_{\textrm{C}}$.
Below $T_{\textrm{C}}$ the temperature dependence of 
the spontaneous magnetization $M_{\textrm{S}}$(T) is given 
by
\begin{equation}
M_{\textrm{S}}(T) = M_{0}({\epsilon})^{-\beta}; {\epsilon} {<} 0\; 
{\textrm{for}}\; H{\rightarrow}0\,
\end{equation}
Above $T_{\textrm{C}}$, the initial susceptibility is given by, 
\begin{equation}
{\chi}_{0}^{-1}(T) = (h_{0}/M_{0}){\epsilon}^{\gamma},
{\epsilon} {>} 0\,
\end{equation}
At $T_{\textrm{C}}$, $M$ and $H$ are related as,
\begin{equation}
M = DH^{1/\delta}, {\epsilon} = 0\,
\end{equation}
Here $M_{0}$, $h_{0}$/M$_{0}$ and $D$ 
are the critical amplitudes.

In the critical region, the magnetic equation of state 
is given by
\begin{equation}
M(H, {\epsilon}) = {\epsilon}^{\beta}f_{\pm}(H/{\epsilon}^{\beta+\gamma})\,
\end{equation}
where
$f_{+}$ for $T{>} T_{\textrm{C}}$ and $f_{-}$ for $T{<}T_{\textrm{C}}$ 
are regular analytic functions. 
Equation (6) implies that $M/{\epsilon}^{\beta}$ 
as a function of $H/{\epsilon}^{\beta+\gamma}$ 
yields two universal curves, one for temperatures 
below $T_{\textrm{C}}$ and the other for temperatures above $T_{\textrm{C}}$.

If a magnetic system is governed 
by various competing couplings and/or randomness,
the properties in the critical region may show
various systematic trends or cross-over phenomena.
In that case, it is useful to generalize 
the power-laws for the critical behavior
by defining effective exponents through
\begin{equation}
\beta_{\textrm{eff}}=\frac{d\;(\ln\, M(\epsilon))}
                          {d\;(\ln\, \epsilon)},\;\;\;\
\gamma_{\textrm{eff}}=-\frac{d\;(\ln\, \chi(\epsilon))}
                           {d\;(\ln\, \epsilon)}\,
\label{Effexps}
\end{equation}
which depend on the separation $\epsilon$ 
from the critical point.\cite{Kouvel64,Dudka03,Dudka05}

In the asymptotic limit $ \epsilon \rightarrow 0$,
the effective exponents approach the universal 
critical (asymptotic) exponent.
The calculations of the logarithmic derivatives 
involves large uncertainties. 
Therefore, we have used two different methods to obtain
estimates on effective exponents:
(i) The complete set of data $\ln\, M(\epsilon)$  or 
$\ln\, \chi(\epsilon)$ vs $\ln\, \epsilon$ is
fitted through a 4th or 5th order polynomial.
The derivatives are calculated analytically which 
gives a rough overview on the dependence of the 
effective exponents on $\epsilon$.
(ii) Further estimates on the effective exponents
were derived from a method via parabolic 
fits, as proposed in Ref.~$[$\onlinecite{Mobius99}$]$,
which is used here for different values of $k=7 \dots 12$ 
neighbouring points. 

\section{Results}\label{Results}
%
\begin{figure*}[tbh]
\begin{minipage}{0.5\linewidth}
\centering
\includegraphics[width=8.0cm]{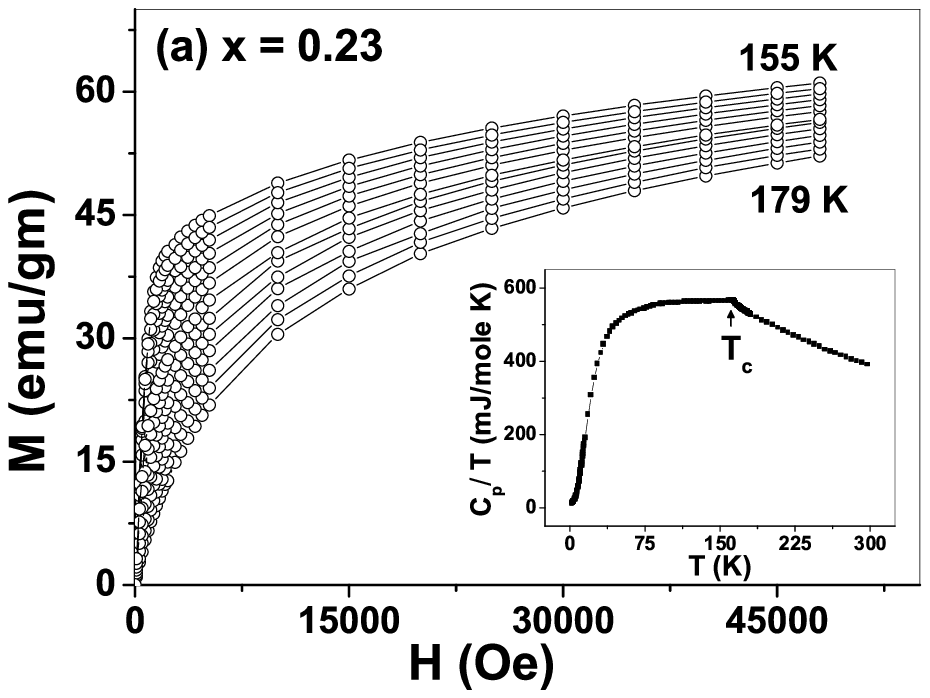}
\end{minipage}%
\begin{minipage}{0.5\linewidth}
\hspace*{0.1cm}
\centering
\includegraphics[width=8.0cm]{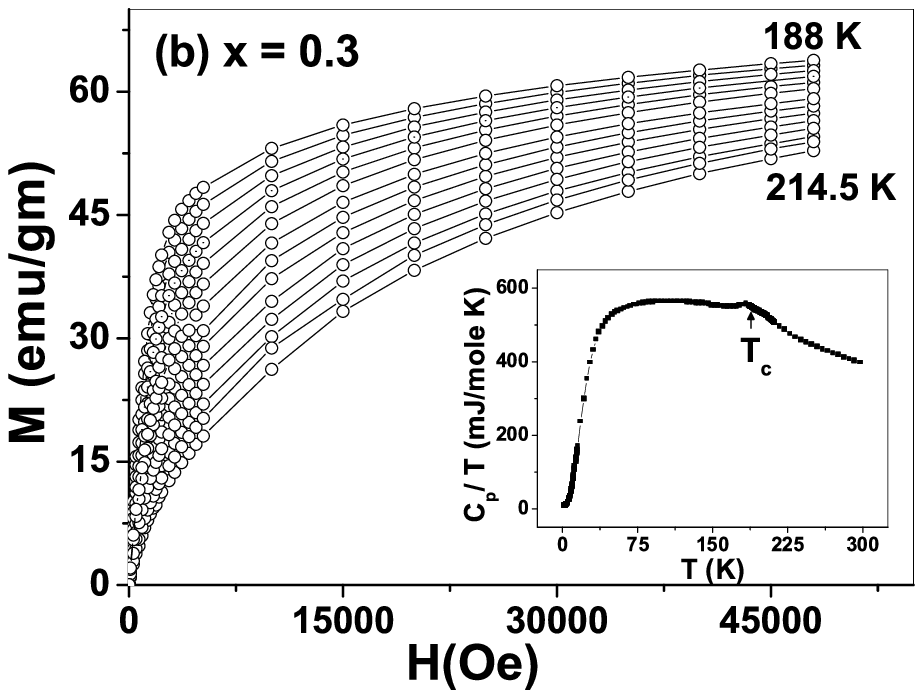}
\end{minipage}
\caption{Magnetization vs. field of Pr$_{1-x}$Pb$_{x}$MnO$_{3}$ 
for (a) $x$ = 0.23 and (b) $x$ = 0.30 from 0 to 5 T 
around $T_{\textrm{C}}$. 
Insets show the specific heat as $C_p/T$ vs. $T$. Weak cusps 
are discernible at $T_{\textrm{C}}$ 
indicating a second order transition.} 
\end{figure*}

Figs.~1(a) and (b) show the $M$ vs. $H$ 
plots for $x=$ 0.23 and 0.3,
respectively, in the temperature range $T_{\textrm{C}}$ ${\pm}$10 K.
Corresponding specific heat data are shown in the 
insets of Figs.~1(a) and 1(b). 
The magnetic transitions are marked by cusps in the specific heat data 
near T$_{\textrm{C}}$. 
This indicates that the transitions are 
likely to be of second order, i.e. continuous
with a negative critical exponent $\alpha < 0$.
%
\begin{figure*}[tbh]
\begin{minipage}{0.5\linewidth}
\centering
\includegraphics[width=8.0cm]{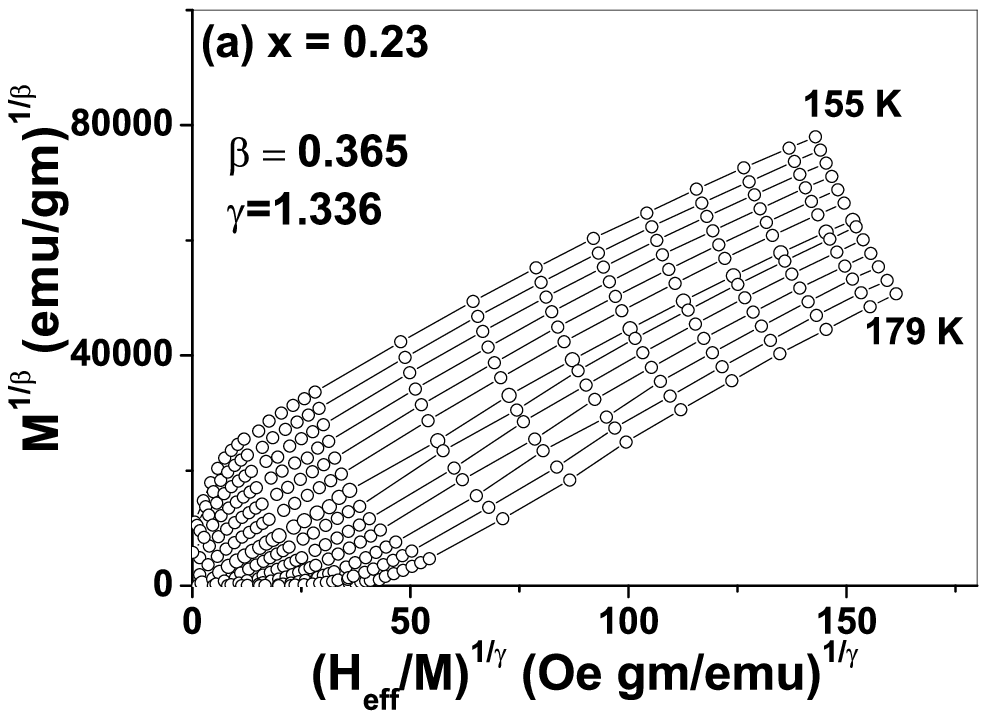}
\end{minipage}%
\begin{minipage}{0.5\linewidth}
\hspace*{0.1cm}
\centering
\includegraphics[width=8.0cm]{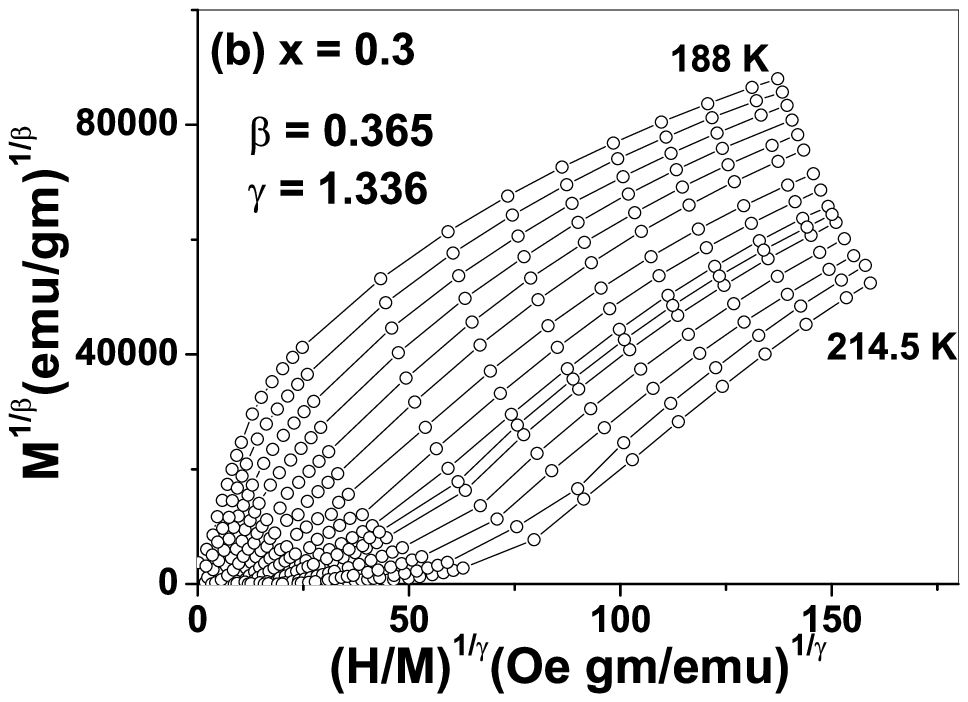}
\end{minipage}
\caption{Modified Arrott plots of Pr$_{1-x}$Pb$_{x}$MnO$_{3}$ for 
(a) $x$ = 0.23 and (b) $x$ = 0.30, 
where ${\beta}$ = 0.365 and ${\gamma}$ = 1.336 
are the trial values
corresponding to 3D Heisenberg ferromagnet.
} 
\end{figure*}

Figs.~2(a) and 2(b) show modified Arrott plots for $M^{1/\beta}$ 
vs. $(H_{\textrm{eff}}/M)^{1/\gamma}$ 
constructed from the $M$ vs. $H$ plots at the different temperatures
by using trial critical exponents $\beta$ and $\gamma$ similar
to those of the 3D Heisenberg magnets.
For the data of $x=0.23$ crystal, this choice produces
a range of nearly parallel linear isothermal curves for higher fields.
This indicates that the critical properties are close to those
of isotropic 3D ferromagnets. 
For the $x=0.30$ crystal, the modified Arrott plots display 
systematic nonlinearities. 
Other choices of trial exponents result in 
stronger deviations from linearity in the modified 
Arrott plots for both crystals.
This has been checked for the values of mean-field, 
tricritical mean-field and 3D Ising-like exponents 
as trial exponents.
Following a standard procedure, the modified Arrott 
plots were extrapolated to the $y$ 
axis intercept corresponding to $M_{\textrm{S}}(T, 0)$.
At $T_{\textrm{C}}$ this extrapolated line passes through the origin. 
The inverse initial susceptibility ${\chi}^{-1}(T)$ 
 is obtained from the ratio of the $y$ intercept 
and the slope above $T_{\textrm{C}}$. 
The $T_{\textrm{C}}$ for $x=$ 0.23 lies between 167~K  and 167.5~K, 
while that for $x=$ 0.30 lies between 197.5~K and 198~K.
For this extrapolation, we have used the values above a field
of about 0.1~T for crystal with $x=0.23$. 
For the crystal with $x=0.3$ roughly linear portions of the 
modified Arrott plot towards high fields have been used,
with more low field values closer to the assumed $T_{\textrm{C}}$.
It is clear, that this approach is a compromise which 
averages out an important physical contributions to the critical
behviour in that case.
The universal scaling laws are almost obeyed with the trial values 
of ${\beta}$ and ${\gamma}$ close to those of the 3D Heisenberg 
universality class for $x$ = 0.30.
%
\begin{figure*}[tbh]
\begin{minipage}{0.5\linewidth}
\centering
\includegraphics[width=8.0cm]{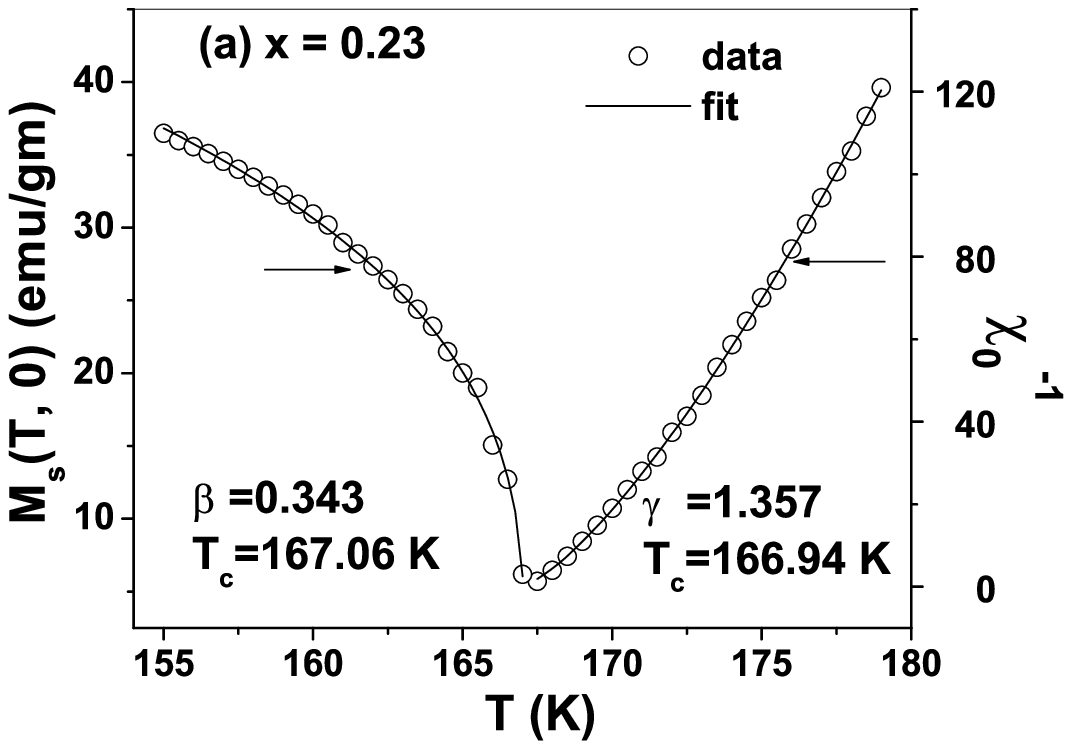}
\end{minipage}%
\begin{minipage}{0.5\linewidth}
\hspace*{0.1cm}
\centering
\includegraphics[width=8.0cm]{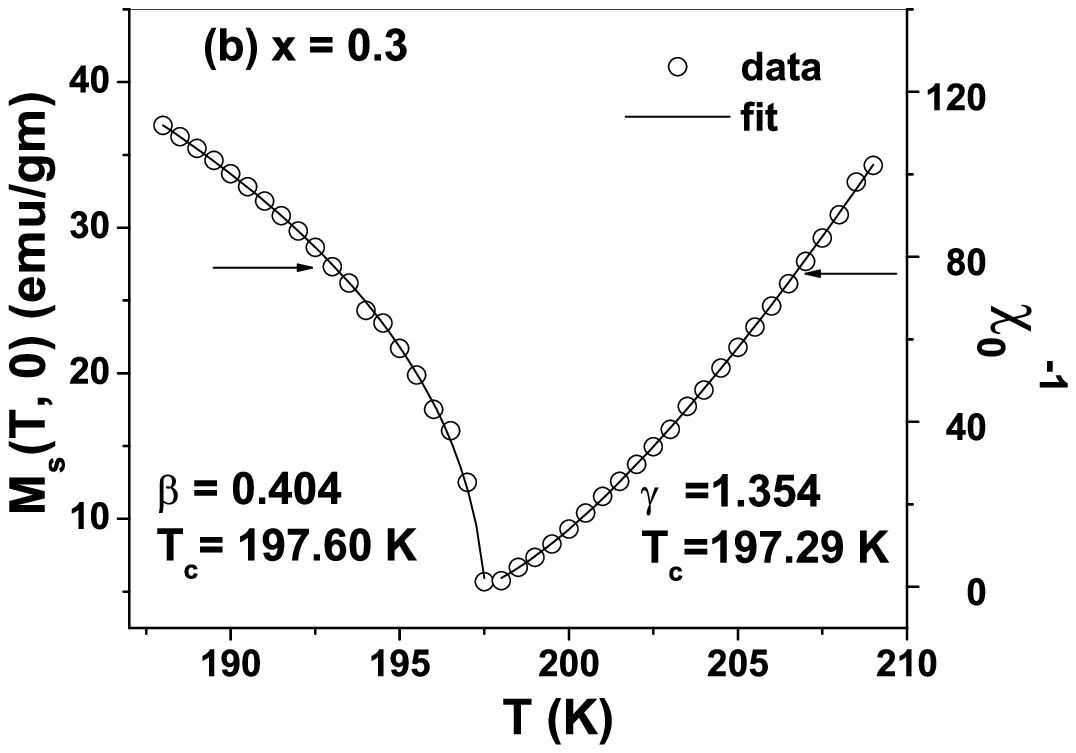}
\end{minipage}
\caption{Spontaneous magnetization (left) and initial 
susceptibility (right) vs. temperature of Pr$_{1-x}$Pb$_{x}$MnO$_{3}$ 
for (a) $x$ = 0.23 and (b) $x$ = 0.30 
to determine ${\beta}$ and ${\gamma}$ respectively.
(For clarity, only part of the $M(H)$ 
data measured around $T_{\textrm{C}}$
are shown in this and next figure.)} 
\end{figure*}

Plots of $M_{\textrm{S}}(T,0)$ vs. $T$ for both compositions 
are shown in Figs.~3(a) and 3(b). 
The values of $T_{\textrm{C}}$ and ${\beta}$ 
were obtained by fitting $M_{\textrm{S}}(T,0)$ vs. $T$ to Eq. (2). 
The ${\chi}^{-1}(T)$ vs. $T$ plots are also 
shown in Figs.~3(a) and 3(b). 
We obtain ${\gamma}$ and $T_{\textrm{C}}$ 
by fitting ${\chi}^{-1}(T)$ to Eq.~(4). 
This procedure is iterated by using 
the derived critical exponents again in a modified Arrott plot. 
Thus, a check has been made that 
the derived critical exponents are close to the trial exponents,
and that the two critical temperatures $T_{\textrm{C}}$ from the fits
are the same within error estimates.
%
\begin{figure*}[tbh]
\begin{minipage}{0.5\linewidth}
\centering
\includegraphics[width=8.0cm]{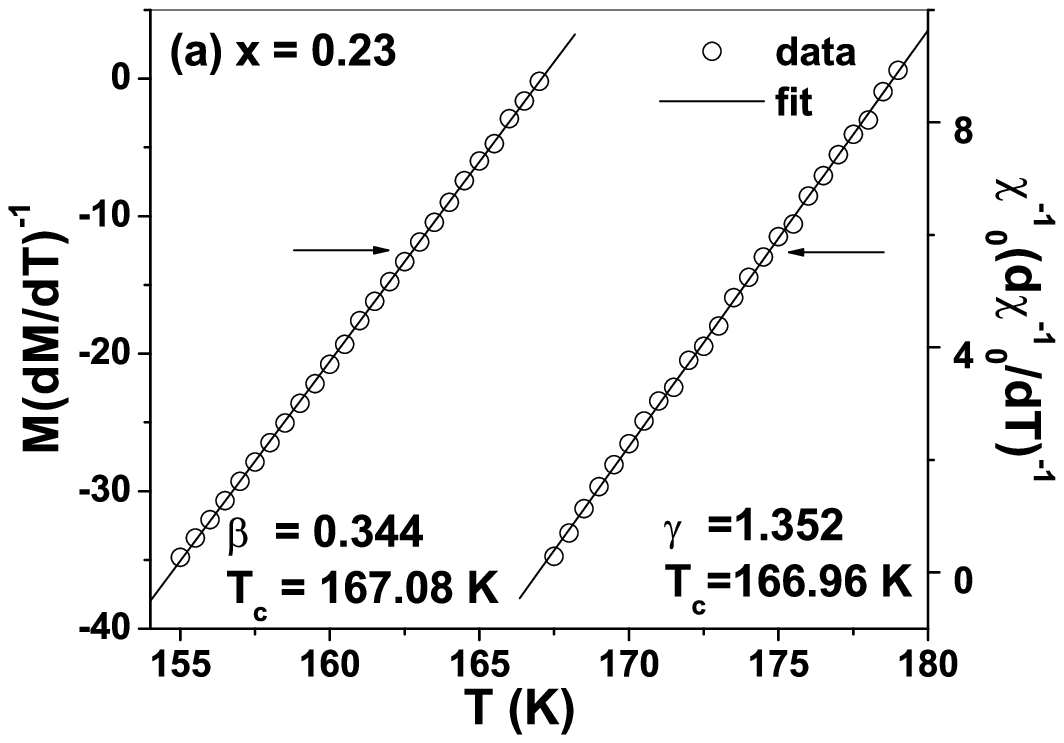}
\end{minipage}%
\begin{minipage}{0.5\linewidth}
\hspace*{0.1cm}
\centering
\includegraphics[width=8.0cm]{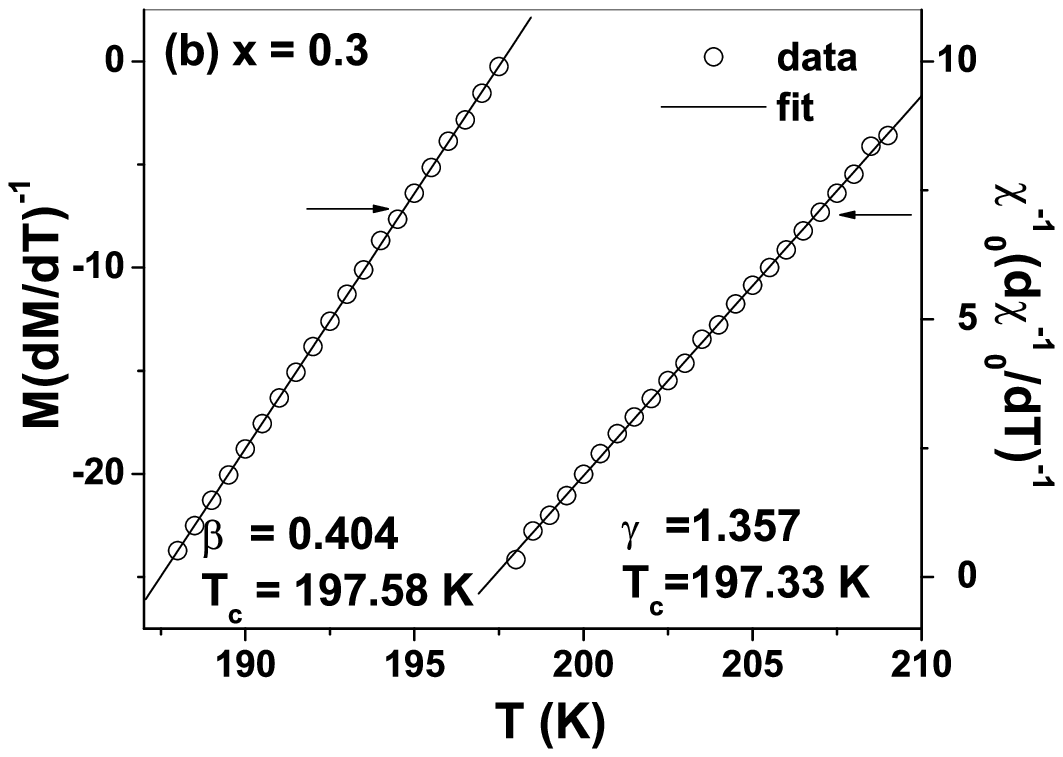}
\end{minipage}
\caption{Kouvel Fisher plots of $M_{S}(T,0)(dM_{S}(T,0)/dT)^{-1}$ 
and ${\chi}^{-1}(T)(d{\chi}^{-1}(T)/dT)^{-1}$ vs. $T$ 
for determination of ${\beta}$ 
and ${\gamma}$ in Pr$_{1-x}$Pb$_{x}$MnO$_{3}$ 
for (a) $x$ = 0.23 and (b) $x$ = 0.30 respectively} 
\end{figure*}

\begin{figure*}[tbh]
\begin{minipage}{0.5\linewidth}
\centering
\includegraphics[width=8.0cm]{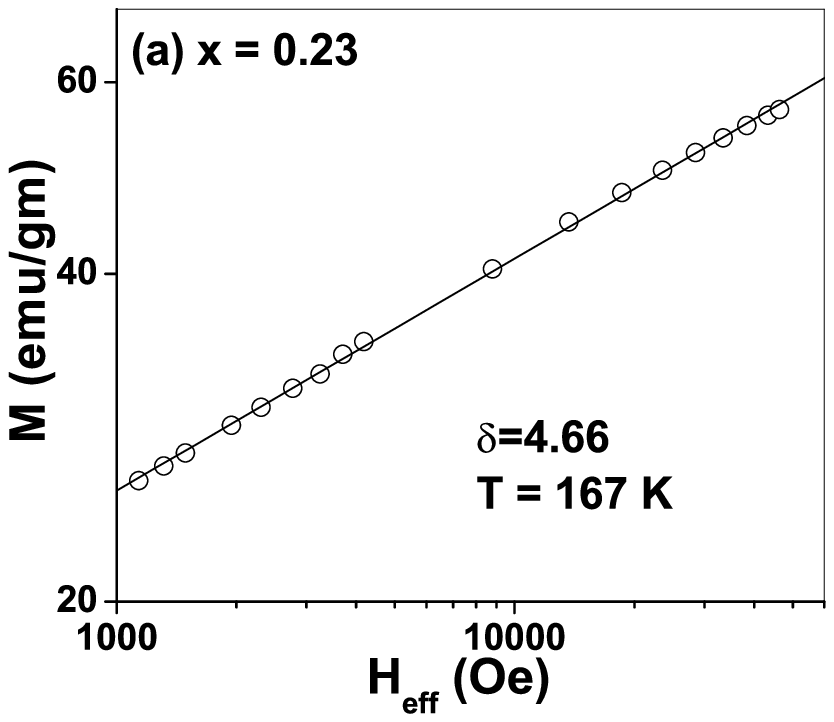}
\end{minipage}%
\begin{minipage}{0.5\linewidth}
\hspace*{0.1cm}
\centering
\includegraphics[width=8.0cm]{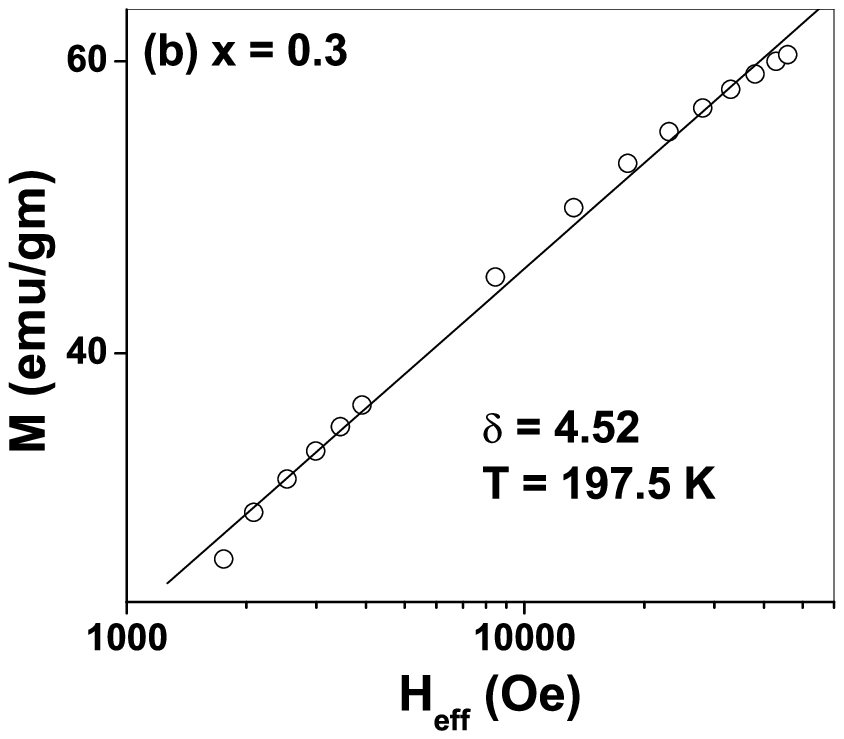}
\end{minipage}
\caption{Critical isotherms for Pr$_{1-x}$Pb$_{x}$MnO$_{3}$ 
for (a) $x$ = 0.23 and (b) $x$ = 0.30 
corresponding to $T_{\textrm{C}}$ = 167 K and $T_{\textrm{C}}$ = 200~K,
respectively.} 
\end{figure*}

\begin{figure*}[tbh]
\begin{minipage}{0.5\linewidth}
\centering
\includegraphics[width=8.0cm]{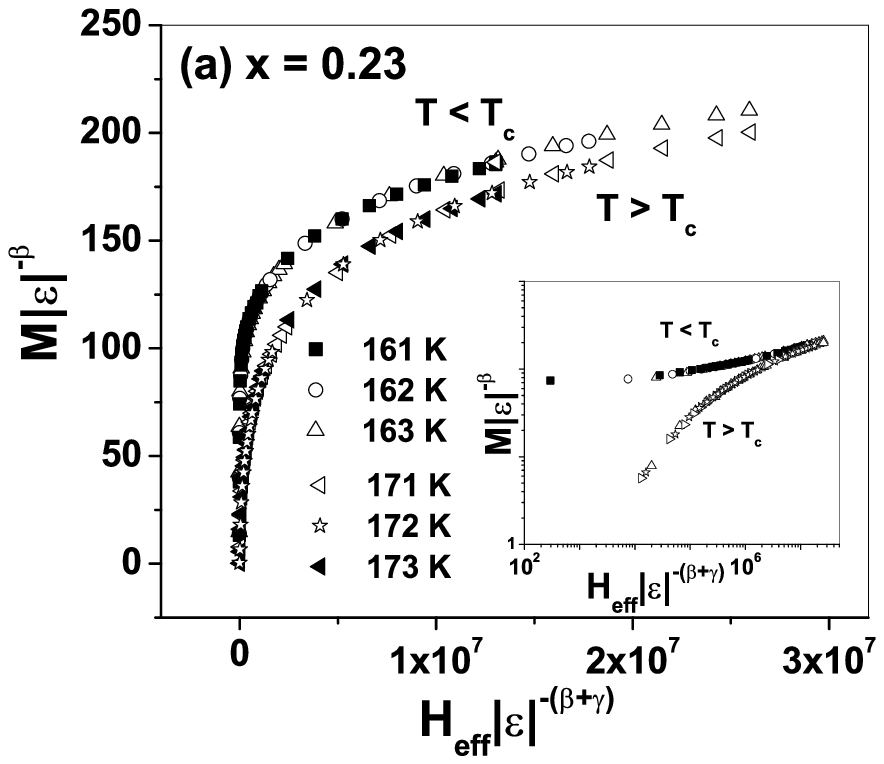}
\end{minipage}%
\begin{minipage}{0.5\linewidth}
\hspace*{0.1cm}
\centering
\includegraphics[width=8.0cm]{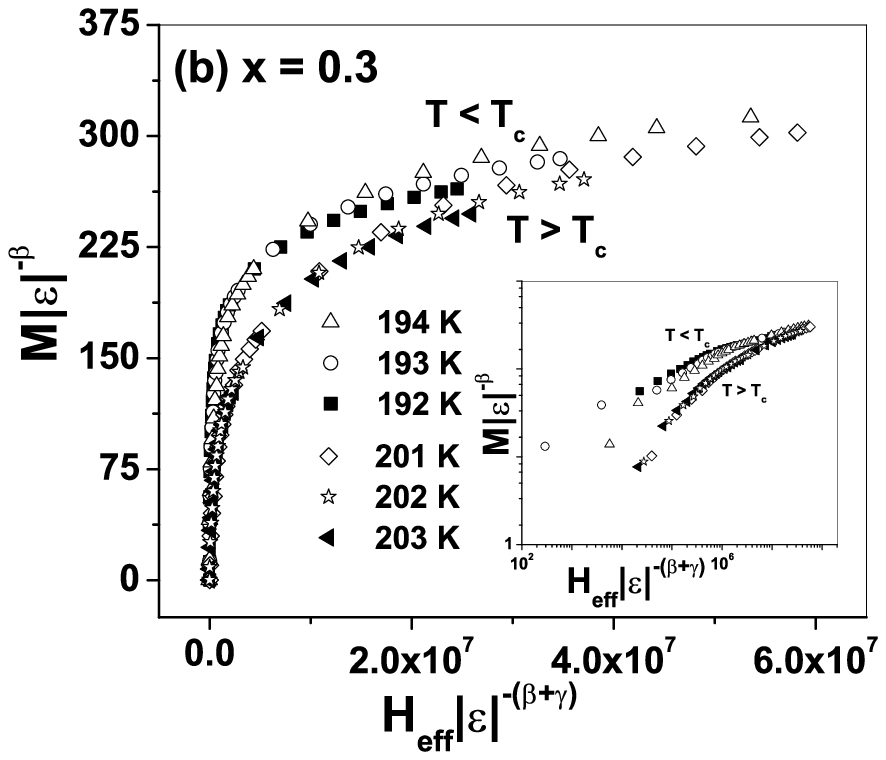}
\end{minipage}
\caption{Scaling plots 
for Pr$_{1-x}$Pb$_{x}$MnO$_{3}$ for (a) $x$ = 0.23 and (b) $x$ = 0.30 
indicating two universal curves below and above $T_{\textrm{C}}$. 
The insets show the plots on a log - log scale.} 
\end{figure*}

The exponents were also obtained by the Kouvel Fisher method,
\cite{Kouvel64}
as shown in Figs.~4(a) and (b) where  $M_{\textrm{S}}(T,0)(dM_{\textrm{S}}(T,0)/dT)^{-1}$ 
is plotted against $T$. 
The plots are straight lines with slope $1/{\beta}$ and $T_{\textrm{C}}$ 
is the ratio of $y$ intercept and slope.
Similarly, in Figs.~4(a) and (b) 
we plot ${\chi}^{-1}(T)(d{\chi}^{-1}(T)/dT)^{-1}$ which shows a linear dependence.
We have listed the critical exponents obtained 
from the modified Arrott plots as well 
as the Kouvel Fisher method 
along with $T_{\textrm{C}}$ in Table~1. 
In Figs.~5(a) and (b), the critical isotherms $M$ vs. $H$ 
for both crystals are plotted, at 167 K and 197.5 K for 
$x=0.23$ and 0.30 respectively. 
At the critical temperature $M$ and $H$ are related by Eq.~(5). 
The high field region of the plot is 
a straight line with slope $1/{\delta}$. 
The value of ${\delta}$ obtained for both compositions are given in Table~1. 
The three exponents derived from our static scaling analysis 
are related by the Widom scaling relation,
\begin{equation}
{\delta} = 1 + {\gamma}/{\beta}\,.
\label{Widom}
\end{equation}
Using this scaling relation and 
the estimated values of $\beta$ and $\gamma$
we obtain ${\delta}$ values which are 
very close to the estimates for $\delta$ from 
the critical isotherms at $T_{\textrm{C}}$.
Thus, the estimates of the critical exponents are consistent.
In the critical region, 
the magnetization and applied field should obey 
the universal scaling behavior. 
In Figs.~6(a) and (b) we show plots of 
$M|{\epsilon}|^{-\beta}$ vs. $H|{\epsilon}|^{-({\beta}+ {\gamma})}$ 
for $x=$ 0.23 and 0.30 respectively.
The two curves represent temperatures below and above $T_{\textrm{C}}$. 
The insets show the same data in log--log scale. 
%

\begin{figure*}[tbh]
\begin{minipage}{0.5\linewidth}
\centering
 \includegraphics[width=8.0cm]{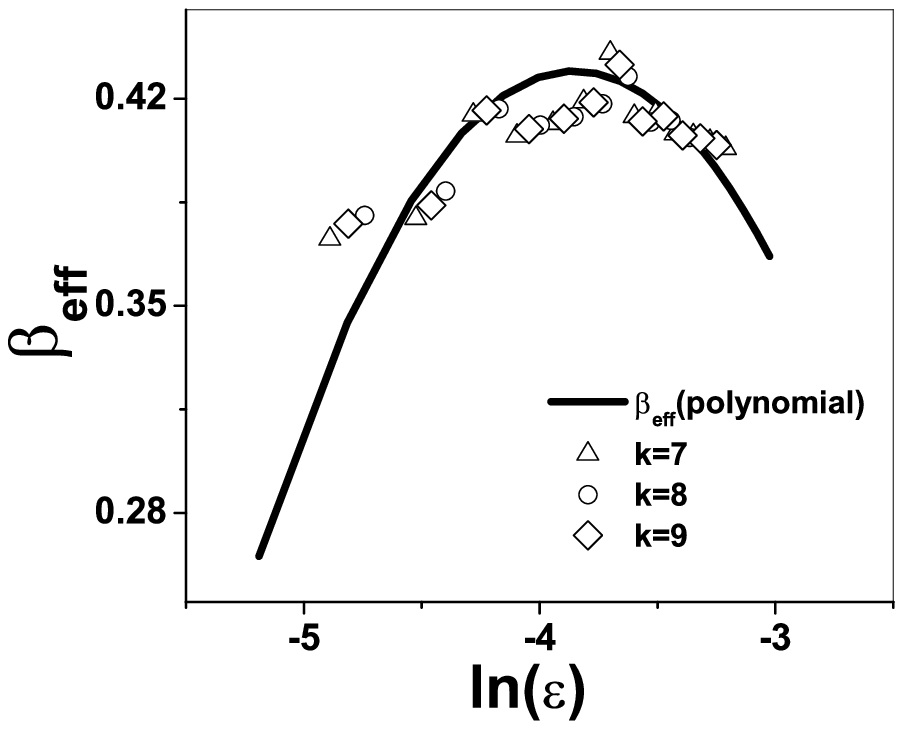}
\end{minipage}%
\begin{minipage}{0.5\linewidth}
\hspace*{0.1cm}
\centering
 \includegraphics[width=8.0cm]{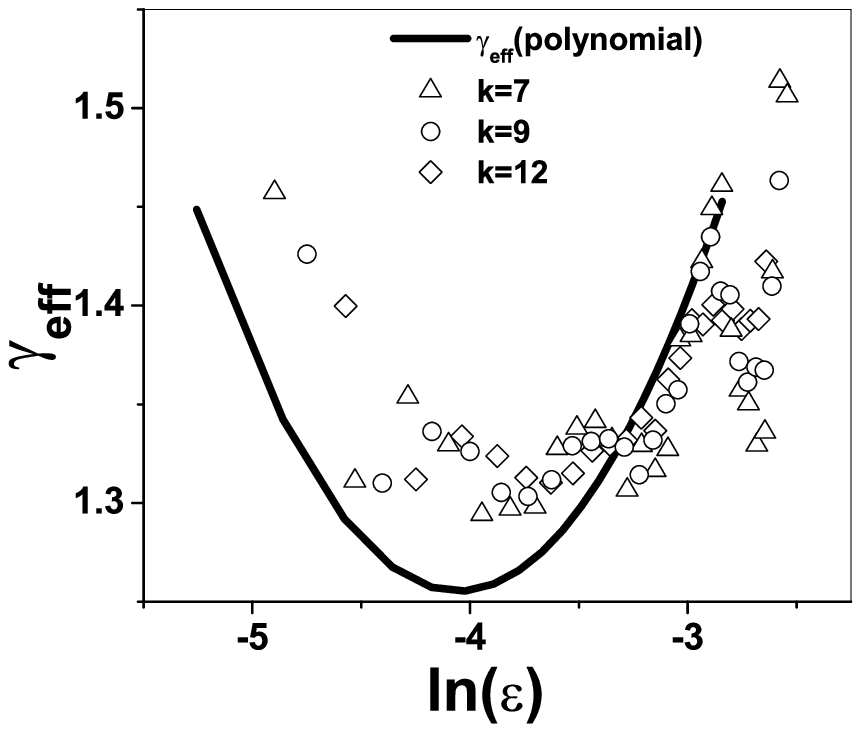}
\end{minipage}
\caption{Effective exponents (a) for static spontaneous magnetization 
$\beta_{\textrm{eff}}$ below $T_{\textrm{C}}$
and (b) for initial susceptibility 
above $T_{\textrm{C}}$ 
vs. reduced temperature $\ln(\epsilon)$ 
for Pr$_{0.70}$Pb$_{0.30}$MnO$_{3}$.
Logarithmic derivatives Eq.~(\ref{Effexps}) 
are from polynomial fits (continuous
thick line) and from a numeric error-improved 
method (Ref.~$[$\onlinecite{Mobius99}$]$)
using $k$ neighbouring data points  (symbols).
}
\end{figure*}
\begin{table*}[tbh]
\caption{Evaluation of critical exponents 
of Pr$_{1-x}$Pb$_{x}$MnO$_{3}$.}
\small{
\begin{center}
\begin{tabular}{|p{2cm}|c|c|c|c|c|c|c|c|c|}\hline 
 & \multicolumn{3}{c|}{} & \multicolumn{3}{c|}{} & \multicolumn{2}{c|}{}\\
Composition & \multicolumn{3}{c|}{Modified Arrott plots } &  \multicolumn{3}{c|}{ Kouvel Fisher method}&\multicolumn{2}{c|}{Critical isotherm}\\ 
 &&&&&&&&\\\hline 
 & ${\beta}$  & ${\gamma}$ & $T_{\textrm{C}}$(average)  & ${\beta}$ & ${\gamma}$ & $T_{\textrm{C}}$(average)  & ${\delta}$(exp) &  ${\delta}$(calc)\\ &&&&&&&&Eq.~(\ref{Widom}) \\\hline
$x$ = 0.23 & 0.343${\pm}$0.005 & 1.357${\pm}$0.020  & 167.00${\pm}$0.13 & 0.344${\pm}$0.001  & 1.352${\pm}$0.006  & 167.02${\pm}$0.04 & 4.69${\pm}$0.02 &4.93\\&&&&&&&&\\\hline 
$x$ = 0.30 & 0.404${\pm}$0.006 & 1.354${\pm}$0.020 & 197.46${\pm}$0.13&0.404${\pm}$0.001 & 1.357${\pm}$0.006 & 197.45${\pm}$0.04  & 4.73${\pm}$0.09&4.37\\ &&&&&&&&\\\hline 
\end{tabular}
\end{center}}
\end{table*}

\section{Discussion}\label{Discussion}
From our analysis, we see that the scaling is 
well obeyed for $x=0.23$.
The value of critical exponents of Pr$_{0.77}$Pb$_{0.23}$MnO$_{3}$ 
are consistent with those of the 3D Heisenberg ferromagnet that has 
critical indices $\beta_H=0.368$, $\gamma_H=1.396$, and $\delta_H=4.80$.
\cite{Campostrini02}
This behavior is quite similar to other CMR manganites belonging 
to the universality class of 
3D isotropic ferromagnets 
with short-range exchange couplings.
\cite{Sahana03,Ghosh06}
The $x$ = 0.23 composition is a ferromagnetic insulator 
at low temperatures. 
However this composition is on 
the crossover region of a MI-transition. 
Around $T_{\textrm{C}}$, there exists 
a small temperature range where it exhibits 
a metallic behavior as mentioned earlier.\cite{Padmanabhan06}
On the other hand, the scaling is poor for the crystal with 
composition $x=0.3$, in particular towards low fields. 
The non-linearity in the modified 
Arrott plot indicate a non-conventional magnetic ordering
behavior in this crystal.
As remarked above, this crystal undergoes 
a metal insulator transition at 235 K. 
This temperature is about 35~K above 
the Curie temperature $T_{\textrm{C}}$,
which is quite unusual for conventional manganites.
\cite{Padmanabhan06,Li05}
One way to interpret the data can be based on the observation
that there are systematic deviations from Curie Weiss behavior 
below 250~K in this compound.\cite{Padmanabhan06}
This may indicate the formation of ferromagnetic clusters,\cite{Li05}
which results in a percolation mechanism for conduction 
and metallic behavior without a magnetic long-range order 
above $T_{\textrm{C}}$.
Then, the non linearity of modified Arrott plots 
in Fig.~3(b) may be attributed 
to the presence of ferromagnetic clusters. 
However, the overall scaling behavior still 
indicates that some form of long-range magnetic 
order is finally established at 
the critical temperature $T_{\textrm{C}}$.
In this case, the clustering behavior 
could be understood as a Griffiths phase 
with randomly varying local 
ordering temperature,\cite{Griffiths69} 
as proposed and described for ferromagnetic
manganites in Refs.~$[$\onlinecite{Salamon02}$]$
and $[$\onlinecite{Deisenhofer05}$]$.
Clustering in the relevant models 
with random-temperature disorder 
for a Heisenberg-like magnet would yield 
complicated cross-over phenomena near criticality.\cite{Dudka03}
In particular, from recent theoretical analysis
for magnetic systems with correlated quenched-disorder,
critical properties with non-universal 
exponents have been found.
For instance long-range correlated quenched-disorder,
as described by the Weinrib-Halperin model 
with correlated random-temperature defects 
falling of as a power law $r^{-a}$ with distance,
can become relevant for Heisenberg-like magnets.\cite{Weinrib83}
In the case of a 3D isotropic Heisenberg-like magnets,
the critical exponents in this random-temperature model 
continuously vary with the exponent $a <3$. 
From a field-theoretical calculation, 
values for $\beta$ in the range 0.387 to 0.384 
and remarkably increasing values of $\gamma$ 
in the range 1.37 to 1.57 have been found.\cite{Prudnikov99}
On the other hand, Heisenberg-like systems 
may show values $\gamma > 1.41$ in 
the presence of extended impurities 
with dimensionality $\varepsilon_d \ge 0.4$.\cite{Blavatska03}
Otherwise, for non-correlated quenched disordered systems
conventional Heisenberg-like critical properties
should be expected. 
Such conventional critical properties have been found 
for the rounded PM--FM transition in La$_{0.67}$Ca$_{0.33}$MnO$_3$ 
with low Ga-substitution on Mn-sites.\cite{Sahana04}
The results on that site diluted manganite 
with full control of the quenched disorder 
demonstrate that the ferromagnetic manganites
do not generically own correlated random-temperature defects, that 
could lead to unconventional critical properties.

If there are correlated random-temperature 
defects in the present Pr$_{0.70}$Pb$_{0.30}$MnO$_3$ 
crystal, their influence must be weak.
Otherwise, the static scaling should hold 
with a consistent, but non-universal 
set of critical exponents, in particular
with an increased value of $\gamma$.
As demonstrated by the extrapolation in the modified 
Arrott plot for relatively large fields and the scaling
plot Fig.~6(b),
we see an effective critical behavior in the crystal $x=0.3$, 
that is roughly consistent with 
the properties of a Heisenberg-like magnet.
Therefore, a clustering leading 
to a remarkable Griffiths phase effect
seems insufficient to explain the anomalous 
magnetic ordering in our system.

The rounding effect, as seen in the 
modified Arrott plot Fig.~2(b), 
is pronounced close to the transition.
As has been pointed out
by Rivadulla et al.\cite{Rivadulla04} 
this behavior is often seen 
in various nominally ferromagnetic manganites.
It indicates a suppression of ferromagnetic order 
by some further coupling effects that are effective 
on longer ranges and may lead to frustration.
In fact, the modified Arrott plots 
resemble those proposed from theoretical considerations
on magnets with frustrated couplings of random-field 
or random-anisotropy type,\cite{Aharony80}
as found, e.g., in experiments 
on rare-earth based disordered magnets.\cite{Gehring90}
The presence of quenched random-fields is difficult
to motivate for an overall ferromagnetic system
in zero external fields, without very particular mechanisms.
Hence, we should expect that the manganites 
behave as random anisotropy Heisenberg-like magnets.
Quenched-random anisotropies 
destroy the long-range magnetic order 
in isotropic magnets.
But, long-range order could be re-established by 
anisotropic distributions of random axes and/or 
by weaker global cubic or uniaxial
anisotropies.\cite{Dudka05}

To discern the influence of such random magnetic couplings,
we have plotted effective criticial exponents
for the crystal with composition $x=0.3$ in Figs.~7(a) 
and (b).
These effective exponents have been derived 
by the logarithmic derivates Eq.~(7) 
for the extrapolated $M_{\textrm{S}}(T, 0)$ and $\chi_0(T)$ 
from the modified Arrott plot. 
As usual, the estimates for $\gamma_{\textrm{eff}}$ and 
$\beta_{\textrm{eff}}$ are prone to large 
uncertainty reflected in the large scatter between 
the different numerical methods used for Fig.~7.
However, there clearly is a systematic non-monotonous shift of 
the exponents with decreasing reduced temperature $\epsilon$. 
In particular, $\gamma_{\textrm{eff}}$
increases with increasing $\epsilon$. 
A maximum or increasing $\gamma_{\textrm{eff}}$ with larger 
$\epsilon$ has similarly been found in a recent investigation 
on a frustrated metallic ferromagnet.\cite{Perumal03} 
Such systematic dependencies of $\gamma_{\textrm{eff}}$ 
are suggested by theoretical investigations 
on effective critical properties
of random anisotropy magnets.\cite{Dudka05}
From our data, we cannot determine 
whether a plateau is reached for the effective 
exponents at small separation $\epsilon$.
It would correspond to the values of the 
asymptotic exponents, if the transition 
is continuous. 
The data may also indicate
a diverging $\gamma_{\textrm{eff}}$ 
and vanishing $\beta_{\textrm{eff}}$.
Such a behavior could indicate 
an infinite susceptibility phase 
or a quasi long-range ordered domain-like state,
without macroscopic spontaneous magnetization 
as proposed for random-anisotropy systems.\cite{Aharony80}
Thus the systematic shift of effective exponents 
suggest the presence of random symmetry breaking couplings 
in the magnetic system of the crystal with $x$ = 0.30. 
Recent investigations on the Griffiths-like properties 
in paramagnetic La$_{1-x}$Sr$_{x}$MnO$_3$ also indicate
that frustrated magnetic couplings may be present
in the ferromagnetic manganites.\cite{Deisenhofer05}
Such additional frustrating couplings explain 
for the Pr$_{0.70}$Pb$_{0.30}$MnO$_3$ crystal
(i) the suppression of the magnetic transition 
temperature $T_{\textrm{C}}$ 
below the MI transition temperature and the occurrence of 
a metal-like conductivity in a paramagnetic state 
(ii) the anomalous rounding as seen in the modified Arrott plot, 
and (iii) the markedly non-monotonous dependence of 
the effective critical exponents near 
the apparent magnetic ordering transition.

\section{Conclusions}\label{Conclusions}

The static scaling analysis near the magnetic
ordering on single crystals of Pr$_{1-x}$Pb$_{x}$MnO$_{3}$ for $x$ = 0.23 
and $x$ = 0.30 has been performed to derive
the critical exponents ${\beta}$, ${\gamma}$, 
and ${\delta}$. 
These exponents were determined by extrapolating
from relatively large fields. 
This means that we probe the system away from criticality at
shorter lengths, where the exchange mechanism
is dominant.
The critical exponents indicate that the underlying 
magnetic system in both the insulating and the 
metallic crystal is similar to that of a conventional
isotropic magnet, as is expected for the manganites.
The exponents are also consistent, as 
the Widom scaling relation between the exponents is obeyed. 
For the insulating crystal with composition $x$ = 0.23 
the data fit well with the universal scaling behavior.
Thus, the magnetic properties in these crystals are 
those of an isotropic magnet with short-range exchange couplings.
They are not subject to remarkable further effects 
as long-range magnetic couplings,
that would lead to non-universal or mean-field properties.
However, we can discern important magnetic coupling effects,
which destroy proper universal scaling 
for the metallic crystal with $x$ = 0.30. 
The scaling analysis as a conventional ferromagnet 
yields exponents, which average out the systematic drift seen 
in the effective exponents. Hence, there is evidence for 
an important cross-over effect in this crystal, 
which intercepts the formation of the conventional magnetic long-range 
order by frustrated magnetic couplings. 

\begin{acknowledgments}
One of the authors(H.L.B.) acknowledges C.S.I.R (India) 
for financial support.
\end{acknowledgments}


 

\section{References}\label{References}

\begin{thebibliography} {99}

\bibitem{Zener51}
C. Zener, Phys. Rev. \textbf{82}, 403 (1951).

\bibitem{Tokura00} 
\textit{Colossal Magnetoresistive Oxides}, 
eds. Y. Tokura (Gordon \& Breach Science Publishers, New York, 2000). 

\bibitem{CNRRao98} 
\textit{Colossal Magnetoresistance, Charge Ordering 
and relative properties of Magnetic oxides}, 
eds. C.N.R.Rao and B.Raveau (World Scientific Singapore, 1998).

\bibitem{Urushibara95} 
A. Urushibara, Y. Moritomo, T. Arima, A. Asamitsu, G. Kido, and Y. Tokura, 
Phys. Rev. B \textbf{51}, 14103 (1995).

\bibitem{Schiffer96} 
P. Schiffer, A.P. Ramirez, W. Bao, and S.-W. Cheong,
Phys. Rev. Lett. \textbf{75}, 3336 (1995).

\bibitem{Martin99} 
C. Martin, A. Maignan, M. Hervieu, and B. Raveau, 
Phys. Rev. B \textbf{60}, 12191 (1999).

\bibitem{Archibald96}
W. Archibald, J. S. Zhou, and J. B. Goodenough, 
Phys. Rev. B \textbf{53}, 14445 (1996).

\bibitem{Motome01} 
Y. Motome and N. Furukawa, 
J. Phys. Soc. Jpn. \textbf{70}, 1487 (2001). 

\bibitem{Lofland97} 
S. E. Lofland, V. Ray, P. H. Kim, 
S. M. Bhagat, M. A. Manheimer, and S. D. Tyagi, 
Phys. Rev. B \textbf{55}, 2749 (1997).

\bibitem{Martin96} 
M. C. Martin, G. Shirane, Y. Endoh, K. Hirota, Y. Moritomo, and Y. Tokura, 
Phys. Rev. B \textbf{53}, 14285 (1996).

\bibitem{Hong01} 
C. S. Hong, W. S. Kim, and N. H. Hur, 
Phys. Rev. B \textbf{63}, 92504 (2001).

\bibitem{Kim02} 
D. Kim, B. Revaz, B. L . Zink, F. Hellman, J. J. Rhyne, 
and J. F. Mitchell,
Phys. Rev. Lett. 89, 227202 (2002).


\bibitem{Padmanabhan06}
B.Padmanabhan, Suja Elizabeth, H. L. Bhat, 
Sahana R{\"o}ssler, K. D{\"o}rr, and K. H. M{\"u}ller, 
J. Magn. Magn. Mater. \textbf{307}, 288 (2006).

\bibitem{Pai03} 
G. V. Pai, S. R. Hassan, H. R. Krishnamurthy, and T. V. Ramakrishnan,
Europhys. Lett. \textbf{64}, 696 (2003).

\bibitem{Morrish69} 
A. H. Morrish, B. J. Evans, J. A. Eaton, and L. K. Leung, 
Can. J. Phys. \textbf{47}  2691 (1969).


\bibitem{Kouvel64}
J. S. Kouvel and M. E. Fisher, 
Phys.  Rev. \textbf{136}, A 1626 (1964).

\bibitem{Dudka03}
M. Dudka, R. Folk, Yu. Holovatch, and D. Ivaneiko,
J. Magn. Magn. Mater. \textbf{256}, 243 (2003).

\bibitem{Dudka05}
M. Dudka, R. Folk, and Yu. Holovatch,
J. Magn. Magn. Mater. \textbf{294}, 305 (2005).

\bibitem{Mobius99}
A. M\"obius,
C. Frenzel, R. Thielsch, 
R. Rosenbaum, C.J. Adkins, M. Schreiber, 
H.-D. Bauer. R. Gr\"otzschel, V. Hoffmann, T. Krieg, N. Matz, 
H. Vinzelberg, and M. Witcomb,
Phys. Rev. B \textbf{60}, 14209 (1999).
%

\bibitem{Campostrini02}
M. Campostrini, M. Hasenbusch, A. Pelissetto, P. Rossi, and E. Vicari, 
Phys. Rev. B \textbf{65}, 144520 (2002).

\bibitem{Sahana03} 
M. Sahana, U. K. R{\"o}ssler, Nilotpal Ghosh, Suja Elizabeth, 
H. L. Bhat, K. D{\"o}rr, D. Eckert, M. Wolf, and K.-H. M\"{u}ller, 
Phys. Rev. B. \textbf{68}, 144408 (2003).

\bibitem{Ghosh06} 
Nilotpal Ghosh, Sahana R{\"o}ssler,  
U. K. R{\"o}ssler, K. Nenkov, Suja Elizabeth, 
H. L. Bhat, K. D{\"o}rr, and K.-H. M{\"u}ller, 
J. Phys.: Condens. Matter \textbf{18}, 557 (2006).

\bibitem{Li05} 
Run-Wei Li, Xin Zhou, Bao-Gen Shen, and Burkhard Hillebrands, 
Phys. Rev. B \textbf{71}, 92407 (2005).


\bibitem{Griffiths69} 
R.B.Griffiths,
Phys. Rev. Lett. \textbf{23}, 17 (1969).

\bibitem{Salamon02}
M.B. Salamon, P. Lin, and S.H. Chun,  
Phys. Rev. Lett. \textbf{88}, 197203 (2002);
M.B. Salamon and S.H. Chun, 
Phys. Rev. \textbf{68}, 014411 (2003).

\bibitem{Deisenhofer05}
J.Deisenhofer, D. Braak, H.-A. Krug von Nidda, 
J. Hemberger, R.M. Eremina, 
V.A. Ivanshin, A.M. Balbashov, 
G. Jug, A. Loidl,
T. Kimura, and Y. Tokura,
Phys. Rev. Lett. \textbf{95}, 257202 (2005).

\bibitem{Weinrib83}
A. Weinbrib and B.I. Halperin,
Phys. Rev. B~\textbf{27}, 413 (1983).

\bibitem{Prudnikov99}
V.V. Prudnikov, P.V. Prudnikov, and A.A. Fedorenko,
J. Phys. A: Math. Gen. \textbf{32}, 8587 (1999).

\bibitem{Blavatska03}
V. Blavats'ka, C. von Ferber, and Yu. Holovatch,
Phys. Rev. B \textbf{67}, 094404 (2003).

\bibitem{Sahana04}
S. R{\"o}ssler, U. K. R{\"o}ssler, K. Nenkov, D. Eckert, 
S.M. Yusuf, K. D{\"o}rr, and K.-H. M\"{u}ller, 
Phys. Rev. B. \textbf{70}, 104417 (2004).

\bibitem{Rivadulla04}
F. Rivadulla, J. Rivas, and J.B. Goodenough,
Phys. Rev. B \textbf{70}, 172410 (2004).
%

\bibitem{Aharony80}
A. Aharony and E.Pytte,
Phys. Rev. Lett. \textbf{45}, 1583 (1980).

\bibitem{Gehring90}
P.M. Gehring, M.B. Salamon, A. del Moral, 
and J.I. Arnaudas,
Phys. Rev. B \textbf{41}, 9134 (1990).

\bibitem{Perumal03}
A. Perumal, V. Srinivas, V.V.Rao, and R.A. Dunlap,
Phys. Rev. Lett. \textbf{91}, 137202 (2003).
%

\end{thebibliography}
\end{document}